\documentclass[12pt]{iopart}

\usepackage{graphicx}

\newcommand{\bean}{\begin{eqnarray}}
\newcommand{\eean}{\end{eqnarray}}
\newcommand{\be}{\begin{equation}}
\newcommand{\ee}{\end{equation}}
\newcommand{\eqref}[1]{(\ref{#1})}

\begin{document}              

\title{Can a dust dominated universe have accelerated expansion?}
\date{\today}

\author{H\aa vard Alnes$^1$, Morad Amarzguioui$^2$ and \O{}yvind
  Gr\o{}n$^{1,3}$}
\address{
  $^1$ Department of Physics, University of Oslo, PO Box 1048
  Blindern, 0316 Oslo, Norway\\
  $^2$ Institute of Theoretical Astrophysics, University of Oslo, PO 
  Box 1029 Blindern, 0315 Oslo, Norway\\
  $^3$ Oslo College, Faculty of Engineering, Cort Adelersgt.~30, 0254 
  Oslo, Norway}

\begin{abstract}
Recently, there has been suggestions that the apparent accelerated
expansion of the universe is due not to a cosmological constant, but
rather to inhomogeneities in the distribution of matter. In this work,
we investigate a specific class of inhomogeneous models that can be
solved analytically, namely the dust-dominated Lema\^itre-Tolman-Bondi
universe models. We show that they do not permit accelerated cosmic
expansion.\\

\noindent \textbf{Keywords:} inhomogeneous universe models -- dark
energy -- accelerated expansion

\end{abstract}

\maketitle

\section{Introduction}

Kolb and coworkers \cite{Kolb:2005me} argued recently that growing
perturbations on a scale larger than the Hubble length may lead to
accelerated expansion as observed from the center of the perturbation.
This was then followed up by Wiltshire \cite{Wiltshire:2005fw} and
Carter et al. \cite{Carter:2005mu}, who considered a dust
dominated universe model where the observed universe is an underdense
bubble in an Einstein-de Sitter universe. They calculated the
luminosity-redshift relationship for their model and found very small
deviations from the standard $\Lambda$CDM model. Thus, they were able
to obtain a good fit to the Hubble diagram of type Ia supernovae. 

However, critical papers have appeared. We will now proceed to give a
brief overview over some of these papers and the arguments therein. By
considering a universe model where sub-Hubble perturbations are
absent, Flanagan \cite{Flanagan:2005} argues that the contributions
from the super-Hubble perturbations to the value of the deceleration
parameter are so small that they cannot be responsible for the
acceleration of the universe. Geshnizjani et al.
\cite{Geshnizjani:2005} argue that to second order in spatial 
gradients the super-Hubble perturbations only amount to a
renormalization of local spatial curvature, and thus cannot account
for the negative deceleration parameter. Making an expansion to 
Newtonian order in potential and velocity, but taking into account
fully non-linear density inhomogeneities, Siegel and Fry
\cite{Siegel:2005} obtain similar results, and conclude that
inhomogeneity contributions cannot mimic the effects of dark energy or
induce an accelerated expansion. 

One of the strongest and most convincing critiques of the conclusions
of Kolb et al. is that of Hirata and Seljak \cite{Hirata:2005}. They
argue from the Raychaudhuri equation that in a dust dominated universe
there must be a non-vanishing vorticity in order to obtain a negative 
deceleration parameter. Then they showed that the perturbations
considered in \cite{Kolb:2005me} have vanishing vorticity and hence
cannot lead to accelerated expansion. In yet another critique of
the work of Kolb et al., R\"as\"anen \cite{Rasanen:2005zy} also claims
that their conclusions are ruled out. The basis for this conclusion
was an analysis of the model proposed in \cite{Kolb:2005me} by
applying the Buchert formalism for backreaction \cite{Buchert:1999er}.
This conclusion is also supported in works by Giovannini
\cite{Giovannini:2005sy,Giovannini:2005ev}, where the author performs
a general analysis of the class of models considered in
\cite{Kolb:2005me}. He finds that dust-dominated models cannot have
accelerated expansion, and hence, the conclusion of Kolb et al. cannot
be correct.

Several of the arguments against accelerated expansion induced by
inhomogeneities in matter are based upon perturbation calculations.
However, as noted in \cite{Geshnizjani:2005}, one way of evading
these arguments is through non-perturbative effects. This is what we
will investigate in the present work. We want to know if Einstein's
field equations permit inhomogeneities to change the positive
deceleration parameter of a dust-dominated homogeneous universe model
to a negative value. 

Unfortunately, a general analysis incorporating all possible
inhomogeneous models is not feasible, so we will instead concentrate
on one specific class of such models, namely the dust-dominated,
spherically symmetric models. Moffat suggested in \cite{Moffat:2005zx}
that it might be possible to find solutions with accelerated expansion.
Considering the field equations for these models, an expression for
the local deceleration parameter was derived, and it was claimed that
this might indeed allow for negative values. Later, in
\cite{Moffat:2005ii}, this claim was moderated considerably. Here it
was claimed that the \emph{volume averaged} deceleration parameter,
and not the local deceleration parameter, could be negative. No
explicit solution of the field equation were actually found in these
analyses. Instead, the conclusions were reached by investigating
the field equations directly.

In this work, we will consider the same class of spherically
symmetric, inhomogeneous universe models and present a general
solution to the field equations. We then show that no such model can
have an accelerated expansion in the sense of a negative local
deceleration parameter.

The structure of this paper is as follows. In Sect.~\ref{sec:ltb} we
present the field equations for a general spherically symmetric,
inhomogeneous universe model. In Sect.~\ref{sec:exprate} we discuss
the properties of these models and whether they can have accelerated
expansion. Finally, in Sect.~\ref{sec:conc} we give a brief summary of
our work and present our conclusion.

\section{Spherically symmetric, inhomogeneous universe models}
\label{sec:ltb}

The spherically symmetric, inhomogeneous universe models are described
by the Lema\^itre-Tolman-Bondi (LTB) space-time
\cite{Lemaitre:1933,Tolman:1934,Bondi:1947}. The line element can be
written as
\be
\label{eq:metric}
ds^2 = -dt^2 + X^2(r,t)dr^2+R^2(r,t)d\Omega^2 \, ,
\ee
where $X(r,t)$ and $R(r,t)$ are general function to be determined by
the field equations and boundary and initial conditions. They can be
thought of as generalized, position dependent scale factors in the
radial and transverse directions. The Einstein field equations are 
\be
G_{\mu\nu} = R_{\mu\nu}-\frac{1}{2}g_{\mu\nu}R = \kappa T_{\mu\nu} \,
,
\ee
where $\kappa = 8\pi G$. 

We assume the universe to be occupied by an ideal fluid with energy
density $\rho(r,t)$ and an isotropic pressure $p(r,t)$. In the
comoving coordinates defined in Eq.~\eqref{eq:metric}, the
energy-momentum tensor of the fluid can be written as 
$T_{\mu\nu}=\mbox{diag}\,(\rho,p,p,p)$. A relation between $X$ and $R$
can be found by solving the $t$-$r$ component of the field equations,
for which the right-hand side vanishes. The relation one finds is 
\be
X(r,t) = \frac{R'(r,t)}{f(r)},
\ee
where $f(r)$ is an arbitrary function of $r$ only. Throughout this
paper, we will use a $' = d/dr$ to denote differentiation with respect
to $r$ and $\dot{ } = d/dt$ for differentiation with respect to $t$.
 
Following Moffat \cite{Moffat:2005zx,Moffat:2005ii}, we define two
"Hubble parameters" 
\be
\label{eq:hubble}
H_\bot = \frac{\dot{R}}{R}\quad\mbox{and}\quad H_r = \frac{\dot{R}'}{R'}\,,
\ee
which are a measure of the local expansion rate in the perpendicular
and radial directions respectively, and the deceleration parameter
\be
\label{eq:5}
q_\bot = -\frac{1}{H_\bot^2}\frac{\ddot{R}}{R} \, .
\ee
Using these definitions, the time-time and space-space components of
the field equations take the form
\be
\label{eq:6}
H_\bot^2 + 2H_r H_\bot - \frac{\beta}{R^2}-\frac{\beta'}{RR'} =
\kappa \rho 
\ee
and
\be 
\label{eq:7}
-6H_\bot^2q_\bot +2H_\bot^2 -2\frac{\beta}{R^2}-2H_r H_\bot +
\frac{\beta'}{RR'} = -\kappa(\rho+3p)\, .
\ee
where $\beta \equiv f^2-1$. These are the generalization of the
Friedmann equations in a homogeneous universe to a spherically
symmetric, inhomogeneous universe. Note that the function $\beta(r)$ is
often written as $\beta(r) = 2E(r)r^2$, where $E(r)$ determines the
local curvature radius.

\section{Expansion rate of dust-dominated LTB models}
\label{sec:exprate}
From now on we will assume the cosmic fluid to be pressure-less
matter. Adding Eqs.~(\ref{eq:6}) and (\ref{eq:7}), we then obtain the 
following expression for the deceleration parameter
\be
\label{eq:23}
q_\bot = \frac{1}{2}-\frac{\beta}{2\dot{R}^2} \,.
\ee
The condition for accelerated expansion now takes the form
\be
\label{acc_cond}
\beta > \dot{R}^2 > 0 \quad \mbox{ or }\quad f^2 > 1+\dot{R}^2 \, .
\ee
Substituting the definition for the deceleration parameter in
Eq.~\eqref{eq:5} into Eq.~\eqref{eq:23}, we arrive at the expression
\be
\label{eq:25}
2R\ddot{R} + \dot{R}^2 = \beta  \, .
\ee
A single integration of this equation yields
\be
\label{eq:26}
R\dot{R}^2 = \beta R+\alpha \quad \mbox{ or }\quad H_\bot^2 =
\frac{\beta}{R^2}+\frac{\alpha}{R^3} \, ,
\ee
where the function $\alpha(r)$ enters as an integration ``constant''
and depends only on the radial coordinate $r$. Comparing this equation
with the ordinary Friedmann equation for homogeneous models, we see
that the dynamical effects of $\beta$ and $\alpha$ are similar to
those of curvature and dust, respectively. Therefore, $\alpha(r)$ is 
regarded as the gravitational mass function, and one often chooses
$\alpha(r) \propto r^3$. Note, however, that $\alpha$ does not appear
in the energy-momentum tensor.

From Eqs.~\eqref{acc_cond} and \eqref{eq:26}, we find that $\alpha$
and $\beta$ must satisfy the inequality 
\be
\label{dec_ie}
-\beta R < \alpha(r) < 0
\ee
in order for the deceleration parameter to be negative. Thus, it
appears that the LTB models seem to allow for accelerated expansion
even for dust-dominated universe models, as long as $\alpha$ and
$\beta$ are chosen appropriately. The dynamical effect of $\alpha$
would then correspond to that of dust with negative energy density in
a homogeneous universe model. It should be noted that the inequality
above forbids accelerated expansion in a ``big bang'' model where the
scale factor has the initial value $R(0,r) = 0$, which implies that
$\alpha(r) >0$. However, this initial condition may not be physically
realistic. The universe may have started with a finite scale factor,
or maybe has collapsed and reached a finite minimum radius before
expanding again. In such models accelerated expansion does not seem to
be forbidden. However, as we will show shortly, such solutions are
unphysical. First, we find a general solution of the field equation
and show how such an accelerated solution behaves.

\begin{figure}[ht!]
\begin{center}
\includegraphics[width=12cm]{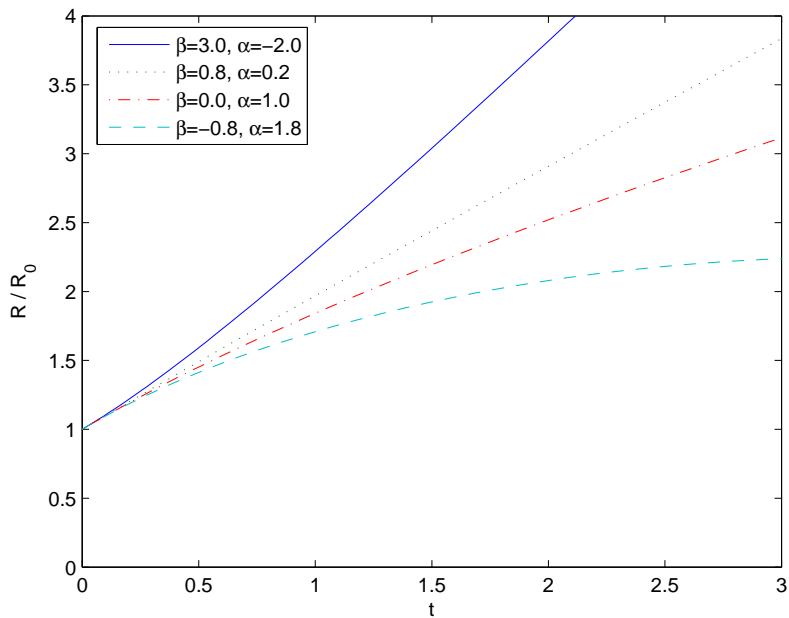}
\caption{The scale factor for some typical values of $\alpha$ and $\beta$}
\label{fig:r-t}
\end{center}
\end{figure}

We introduce a conformal time $\eta$ defined via the differential
relation $\beta^{1/2}dt = Rd\eta$. This allow us to integrate
Eq.~(\ref{eq:26}) to give a parametric solution in terms of $\eta$.
The solutions can be grouped into three different classes according to
the sign of the local curvature $\beta$. Choosing $t_0 = \eta_0 = 0$,
the three classes are
\bean
\label{eq:32A}
R &=& \frac{\alpha}{2\beta}(\cosh \eta - 1)+R_0\left[\cosh \eta +
  \sqrt{\frac{\alpha+\beta R_0}{\beta R_0}}\sinh \eta\right]\\ 
\sqrt{\beta}t &=& \frac{\alpha}{2\beta}(\sinh \eta-
\eta)+R_0\left[\sinh \eta + \sqrt{\frac{\alpha+\beta R_0}{\beta
      R_0}}\left(\cosh \eta -1\right)\right]\quad \mbox{, } \beta >
0\nonumber 
\eean
for positive $\beta$,
\bean
\label{eq:32B}
R &=& \frac{\alpha}{2|\beta|}(1-\cos \eta)+R_0\left[\cos \eta
  +\sqrt{\frac{\alpha+\beta R_0}{|\beta|R_0}}\sin \eta\right]\\ 
\sqrt{|\beta|}t &=& \frac{\alpha}{2|\beta|}(\eta-\sin \eta)+R_0
\left[\sin \eta +\sqrt{\frac{\alpha+\beta R_0}{|\beta|R_0}}(1-\cos
  \eta)\right]\quad \mbox{, } \beta<0\nonumber
\eean
for negative $\beta$, and finally
\be
\label{eq:32C}
R = \left(R_0^{3/2}+\frac{3}{2}\sqrt{\alpha}t\right)^{2/3} 
\ee
for vanishing $\beta$. We allow the initial size of the universe to be
non-zero, hence the non-vanishing value for $R_0\equiv R(t=0,r)$.
Eqs.~(\ref{eq:32A})-(\ref{eq:32C}) with $R_0 = 0$ represent the usual
form of the LTB solution of the field equations. The scale factor $R$
as a function of time is shown for some typical parameters in
Fig.~\ref{fig:r-t}. The top curve represents a solution with
accelerating expansion.

In order for a dust-dominated solution in the LTB space-time to
undergo an epoch of accelerated expansion, the inequality in
Eq.~\eqref{dec_ie} has to be satisfied. Can a physically realistic
solution satisfy this?  Differentiating Eq.~(\ref{eq:26}) and comparing
the result with Eq.~(\ref{eq:6}) we obtain a relation between the
density distribution and $\alpha$ for this class of models,
\be
\label{eq:33}
\kappa \rho = \frac{\alpha'}{R^2 R'} = 3\frac{\alpha'}{V'}\,,
\ee
where $V=R^3$ is a comoving volume. A physically realistic model must
have positive density everywhere, and hence, it must have $\alpha' >
0$. Furthermore, the angular part of the line element is
$R^2d\Omega^2$ where $d\Omega$ is a solid angle element. At the
origin, $r=0$, this part of the line element must vanish, and thus
$R(0,t)=0$. From Eq.~(\ref{eq:26}) we then get that $\alpha(0) = 0$.
Since $\alpha'(r)>0$, it follows that $\alpha(r) \ge 0$ for all $r$.
But, as we showed above, accelerated expansion is only possible for
models with $\alpha(r)<0$. Hence, the dust dominated LTB universe
models must have decelerated expansion. 

We shall now arrive at the same conclusion by an alternative
approach related to the Raychaudhuri equation, which describes
the flow of a cosmological fluid in space-time. It can be written as
\begin{equation} \label{def:Ray}
  \nabla_{\mu}a^{\mu}=\dot{\theta}+\frac{\theta^2}{3}+2(\sigma^2-\omega^2) 
  + R_{\mu\nu}u^{\mu}u^{\nu}\,,
\end{equation}
where $a^{\mu}\equiv u^{\mu}_{;\alpha}u^{\alpha}$ is the four-acceleration
of the fluid, $\theta\equiv u^{\mu}_{;\mu}$ is the expansion rate of
the fluid and $\sigma$ and $\omega$ are the shear and vorticity
scalars, respectively. The latter two quantities are defined as
\begin{equation}
  \sigma^2=\frac{1}{2}\sigma_{\mu\nu}\sigma^{\mu\nu}\quad\mbox{and}\quad 
  \omega^2=\frac{1}{2}\omega_{\mu\nu}\omega^{\mu\nu}
\end{equation}
where
\begin{equation} \label{sheartensor}
  \sigma_{\mu\nu}=u_{(\mu;\nu)}+a_{(\mu}u_{\nu)}-\frac{1}{3}\theta
  (g_{\mu\nu}+u_{\mu}u_{\nu})
\end{equation}
and
\begin{equation} \label{vortensor}
  \omega_{\mu\nu}=u_{[\mu;\nu]}+a_{[\mu}u_{\nu]}\,.
\end{equation}
In a comoving reference frame the four-velocity of the fluid can be
written as $u^{\mu}=[1,0,0,0]$. The divergence of the
four-acceleration will then vanish, and the Raychaudhuri equation can 
be simplified to
\begin{equation}
\label{eq:rayeq}
  0=\dot{\theta}+\frac{\theta^2}{3}+2(\sigma^2-\omega^2)+ R_{00}\,.
\end{equation}
Analogously to the Hubble parameter in FRW models, we can define an 
effective local Hubble parameter from the expansion rate of the fluid:
\be
H=\frac{1}{3}u^\mu_{;\mu}\,.
\ee
This allow us to define an effective deceleration parameter in terms
of $H$ and $\dot{H}$:
\be
q\equiv-\frac{\dot{H}+H^2}{H^2}=-1-3\dot{\theta}/\theta^2\,.
\ee
Using the relation $R_{00}=\frac{\kappa}{2}(\rho+3p)$,
Eq.~(\ref{eq:rayeq}) can be written
\begin{equation}
\label{raycom}
  \theta^2q=6(\sigma^2-\omega^2)+\frac{3}{2}\kappa(\rho+3p)\,.
\end{equation}
While Eq.~(\ref{raycom}) is valid for any metric, let us now
specialize to the case of an LTB model. In \cite{Moffat:2005zx} it was
claimed that the vorticity scalar is non-vanishing for the models
considered in the paper, which may permit negative values of q. We
find, however, that the vorticity tensor~(\ref{vortensor}) vanishes
for a comoving fluid in the metric~(\ref{eq:metric}). This
incorporates also the models considered in \cite{Moffat:2005zx}. For
this metric the shear can be written as  
\begin{equation} \label{shearval}
  \sigma^2=\frac{1}{3}(H_r-H_{\perp})^2\,.
\end{equation}
Finally, assuming the fluid to be pressure-less, Eq.~(\ref{raycom})
reduces to
\begin{equation}
  \theta^2 q=2(H_r-H_{\perp})^2+\frac{3}{2}\kappa\rho\,.
\end{equation}
This equation shows that the dust dominated LTB-models have $q \ge 0$,
meaning that they cannot have accelerated expansion.

Although we have now shown that the LTB models cannot have accelerated
expansion, this doesn't necessarily mean that they cannot explain the
data used to infer this expansion. The first data used to conclude
that the expansion seems to be accelerating were the measurements of
the luminosities of supernovae of type Ia (SNIa)
\cite{riess98,perlmutter99}. Since then newer data have appeared that
strengthen this claim even further
\cite{tonry03,knop03,Riess:2004nr}. However, accelerated expansion 
follows logically from these data only if one assumes that the
universe is homogeneous. The added freedom of having a position
dependent expansion in LTB models allows one to explain the data
without the need for the expansion to accelerate locally. The
explanation would then be that the expansion rate is highest at $r=0$
and decreases with distance from the center, since the oldest
supernovae are also farthest away. In this case a decelerating
universe would seem to be accelerating. C\'el\'erier shows in
\cite{celerier:99} how such inhomogeneities can mimic the effects of
dark energy in supernova observations, when interpreted within the
framework of FRW models. More specifically, Iguchi et al.
\cite{Iguchi:2002} show that models with $\alpha(r)>0$ can reproduce
the luminosity distance-redshift relationship of a $\Lambda$CDM model
up to $z \sim 1$, but not for higher redshifts. However, when compared
to actual SNIa observations, it is easy to find models that fit the
data even better than the best-fit $\Lambda$CDM model
\cite{alnes:2005}. In fact, as we show in \cite{alnes:2005},  
these models might also be compatible with the CMB power spectrum,
even though they do not contain any form of dark energy.

Finally, we wish to stress the point that in this work we have shown that 
the LTB models cannot have an expansion that is \emph{locally} accelerated. 
This does not  exclude the possibility that there can be a so-called 
\emph{volume averaged acceleration},  where a scale factor defined via the 
physical volume of a comoving region  is found to have a positive double time 
derivative. Recent papers discussing the  possibilities of this effect 
explaining the apparent accelerated expansion are Refs.~\cite{Kolb:2005da, 
Buchert:2005xf, Buchert:2005kj, Buchert:2006ya, Nambu:2005zn, Ishibashi:2005, 
Mansouri:2005rf, Mansouri:2006ua, Kasai:2006bt, Paranjape:2006cd, Parry:2006uu, 
Kai:2006ws}. At the present time there does not appear to exist a consensus 
among cosmologists on this issue.

\section{Conclusion}
\label{sec:conc}
Recently it was claimed by Kolb et al. \cite{Kolb:2005me} that
inhomogeneous perturbations of FRW universe models can result in
accelerated cosmic expansion, and thereby eliminate the need to
postulate the existence of the mysterious dark energy. However,
several papers have appeared since criticizing this work heavily.
These attribute the effects claimed by Kolb et al. to incorrect
handling of second order terms in the perturbative expansion. When
dealt with correctly, one would then assume that these terms do not
lead to accelerated expansion. However, this leaves open the
possibility that the full, non-perturbative solutions of the Einstein
equations for inhomogeneous models might exhibit accelerated expansion.

In this work we investigated a special class of inhomogeneous models
which can be solved analytically, the so-called
Lema\^itre-Tolman-Bondi (LTB) models. These are the spherically
symmetric, inhomogeneous models. We found a general solution to a
dust-dominated LTB model, and showed that there exist solutions that
appear to have accelerated expansion. However, these solutions turned
out to be unphysical. Finally, it was shown that any physically realistic
solution of Einstein's equations must have decelerated expansion if it
contains only matter.

\section*{Acknowledgments}
MA acknowledges support from the Norwegian Research Council through
the project 159637/V30 -- ``Shedding Light on Dark Energy''.
\section*{References}
\bibliography{dust_iop}

\end{document}